\documentclass[twocolumn,letter]{jpsj3}
\usepackage{txfonts}
\usepackage{graphicx}
\usepackage{dcolumn}
\usepackage{bm}
\usepackage{color}

\def\runauthor{Yu {\sc Itino} Takahiro {\sc Ohkuma} and Takao {\sc Ohta}}

\renewcommand{\thepage}{}
\newcommand{\del}{\partial}
\newcommand{\abs}[1]{\left\lvert#1\right\rvert}
\newcommand{\avg}[1]{\left\langle#1\right\rangle}

\title{Collective Dynamics of Deformable Self-Propelled Particles \\
with Repulsive Interaction}
\author{Yu {\sc Itino}, Takahiro {\sc Ohkuma} and Takao {\sc Ohta} \footnote{E-mail:
takao@scphys.kyoto-u.ac.jp} }
\inst{Department of Physics, Graduate School of Science, Kyoto
University, Kyoto 606-8502, Japan}

\date{\today}

\abst{We investigate dynamics of  deformable self-propelled particles with a repulsive interaction whose
magnitude depends on the relative direction of elongation of  a pair of particles. A collective motion of the 
particles appears in two dimensions. However this ordered state becomes unstable when the particle density exceeds a certain 
critical threshold and the dynamics becomes disorder.  We show by a mean field analysis that this novel transition characteristic to deformability occurs due to a saddle-node bifurcation.}
\kword{self-propelled particle, collective dynamics, saddle-node bifurcation.}

\begin{document}
\maketitle

Dynamics of self-propelled objects have been studied for these more than one decade. Various collective motions have been found for interacting particles \cite{Vicsek, Toner, Sano} where the hydrodynamic formulation has been developed \cite{Baskaran, Ramaswamy}. See ref. \cite{Ebbens} for a review of experimental studies. 
Effects of noises have also been investigated in various model systems to find a rich variety of collective dynamics \cite{Chate04, Mikhailov05, Perunani, Teeffelen, Chate10}.
However, almost all of the previous studies assume that objects are rigid despite the fact that coupling between shape deformation and migration in self-propulsion is common in biological microorganisms \cite{Cox, Maeda08} . 
In the present paper, we are concerned with the deformation effect on the collective dynamics of self-propelled particles.

Quite recently, we have introduced the model system where the objects are deformable  \cite{Ohta09}. We have shown that deformation produces a lot of interesting dynamics of individual particles such as rotating motion, zig-zag motion,  chaotic motion and helical motion \cite{Ohta09, OOS, Hiraiwa10-1, Hiraiwa10-2}. These results clearly indicate that deformation is relevant to the dynamics. In the present paper, we study dynamics of assembly of
deformable self-propelled particles where there is a repulsive short range interaction between a pair of particles such that the magnitude depends on the relative direction of elongation. Our main concern is  stability of collective dynamics due to deformation of individual particles.

We consider interacting particles in two dimensions.  The time-evolution of the {\it i}-th particle obeys the following set of equations for the position of the center of mass $r^{(i)}_{\alpha}$, the velocity $v^{(i)}_{\alpha}$ and the deformation tensor $S^{(i)}_{\alpha\beta}$
\begin{eqnarray}
 \frac{d}{dt}r^{(i)}_{\alpha} &=& v^{(i)}_{\alpha} ,
 \label{r} \\
 \frac{d}{dt}v^{(i)}_{\alpha} &=& \gamma v^{(i)}_{\alpha} - \abs{\bm{v}^{(i)}}^{2}v^{(i)}_{\alpha} - aS^{(i)}_{\alpha\beta}v^{(i)}_{\beta} + f^{(i)}_{\alpha} ,
 \label{v} \\
 \frac{d}{dt}S^{(i)}_{\alpha\beta} &=& -\kappa S^{(i)}_{\alpha\beta} + b\left(v^{(i)}_{\alpha}v^{(i)}_{\beta}-\frac{1}{2}\abs{\bm{v}^{(i)}}^{2}\delta_{\alpha\beta}\right) ,
 \label{S}
\end{eqnarray}
where $\gamma \ge 0$, $\kappa \ge 0$, $a$ and $b$ are constants.  This  set of equations is an extension of a single particle case \cite{Ohta09}. The tensor $S$ is represented in terms of the unit normal  $\bm{n}$ parallel to the long axis of an elongated particle as
\begin{equation}
 S^{(i)}_{\alpha\beta} = s_{i}\left(n^{(i)}_{\alpha}n^{(i)}_{\beta}-\frac{1}{2}\delta_{\alpha\beta}\right) ,
 \label{defS}
\end{equation}
with $s_{i}$ the magnitude.
The last term in eq. (\ref{v}) stands for the interaction between particles and is given by
\begin{equation}
 \bm{f}^{(i)} = K\sum\limits_{j=1}^N\bm{F}_{ij}Q_{ij} ,
 \label{f}
\end{equation}
where $K$ is a positive constant, $N$ the total number of particles and 
$ \bm{F}_{ij}= -\del U(\bm{r}_{ij})/\del\bm{r}_{ij}$ with $\bm{r}_{ij}=\bm{r}_{i}-\bm{r}_{j}$.
The form of the potential  $U(\bm{r}_{ij})$ will be specified later. The other factor $Q_{ij}$ expresses the dependence of the relative angle of a pair of elongated particles such that when the interacting two particles are parallel, the repulsive interaction is weaker. It is defined by
\begin{eqnarray}
 Q_{ij}= 1+\frac{Q}{2}\mathrm{tr}(S^{(i)}-S^{(j)})^{2} ,
\label{Q}
\end{eqnarray}
where $Q$ is a positive parameter. 

We have introduced the force (\ref{f}) to take into consideration, approximately, of the excluded volume effect of elongated particles. Since the force depends on the variable $S_{\alpha \beta}^{(i)}$, it is possible generally that there is a counter term  in eq. (\ref{S}). However, for simplicity, such a contribution is not considered in the present study.

Equation  (\ref{S}) implies that if the velocity is zero, $v_{\alpha}^{(i)}=0$, the particle does not deform and takes a circular shape in two dimensions.
One of the characteristic features of the set of equations (\ref{r}), (\ref{v}) and (\ref{S}) is that, when the interaction term is absent,  these equations do not contain the particle size explicitly as a parameter. Therefore the interaction range in the potential $U$ is the only fundamental length scale of the system.

We introduce an orientational order parameter
 defined by
\begin{equation}
\Phi = \abs{\frac{1}{N}\sum\limits_{j=1}^{N}e^{2i\theta_{j}}}  ,
\label{Phi}
\end{equation}
where  $\theta_j$ is defined through the relation $\bm{n}^{(i)}=(\cos\theta_{i}, \sin\theta_{i})$.
In order to quantify the degree of coherent motion,  we may define an alternative order parameter \cite{Vicsek}
\begin{equation}
\Phi_v =  \frac{1}{N}\abs{\sum\limits_{j =1}^{N}\frac{\bm{v}^{(j)}}{\abs{\bm{v}^{(j)}}}}= \abs{\frac{1}{N}\sum\limits_{j =1}^{N}e^{i\phi_{j}}} ,
\label{Phi2}
\end{equation}
where we have put $\bm{v}^{(i)}=\abs{\bm{v}^{(i)}}(\cos\phi_{i}, \sin\phi_{i})$. In the following, we shall use $\Phi$ but we have verified numerically that the difference between $\Phi$ and $\Phi_v$ is negligibly small in the plots in Figs.  \ref{fig2} and  \ref{fig3} below.

\begin{figure}[t]
\centering
  \includegraphics[width=0.8\hsize]{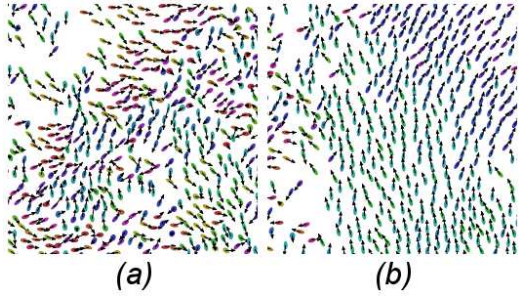}
  \caption{(Color online) Snapshots of the time-evolution of particles at (a) $t=256$ and (b) $t=512$ for $\rho=0.075$ and $N=8192$ starting with a random initial configuration.  These figures display only the  $1/4\times 1/4$ area of the system.   The colors distinguish the direction of elongation and the arrows indicate the direction of migration. }
  \label{fig1}
\end{figure}

\begin{figure}[t]
\centering
  \includegraphics[width=0.9\hsize]{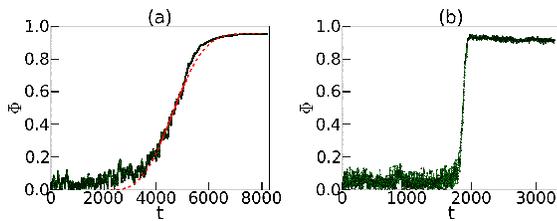}
  \caption{(Color online) Growth of order for (a) $\rho=0.0032$ and  (b) $\rho=0.0852$, and  for $N=512$.  The curve given by eq. (\ref{avrami}) is shown by the broken line in Fig. (a).}
  \label{fig2}
\end{figure}

We shall solve eqs.   (\ref{r}), (\ref{v}) and (\ref{S}) numerically. 
 Hereafter we show the results for the potential
\begin{equation}
U(\bm{r}_{ij})=\exp\left(-\frac{r^{2}_{ij}}{2\sigma^{2}}\right) .
\label{potential}
\end{equation}
The parameters are chosen as  $a=1$ and $b=0.5$ so that each particle tends to elongate to the direction parallel to the velocity \cite{Ohta09}.
We fix the parameters $\gamma=\kappa=1$ in which the particle undergoes straight motion \cite{Ohta09}.
Unless stated explicitly, other parameters are  chosen as $\sigma=1$, $K=5$ and $Q=50$. The set of equations is solved by the Euler explicit method with the time increment  $\delta t=1/64$.  Particles are confined in a square area  with size $L$ and the periodic boundary conditions are imposed.  In most of the simulations, we choose  $N=$512, but the cases  $N=$1024, 2048 and 8192 are also examined to check the number dependence. The density $\rho = N/L^{2}$ is varied by changing the system size $L$. 

Appearance of the collective motion of self-propelled particles is shown in Fig.   \ref{fig1} starting from a disordered initial condition. The particles constitute locally an approximate hexagonal lattice such that  one of the fundamental reciprocal lattice vectors is parallel to the propagating direction. Another orientation of the lattice such that one of the primitive lattice vectors is parallel to the propagating velocity has not been observed. We note that the propagating lattice is not completely stationary. Each particle exhibits a small  oscillation (or fluctuation) in its center of mass perpendicularly to the propagating direction. 

Figure \ref{fig2} shows the growth of the order parameter $\Phi$ for the dilute regime $\rho=0.0032$ and the intermediate  regime $\rho=0.0852$ obtained  for $N=512$. Initially the velocity and the position of the individual particles are random.  When the density is low, the order parameter starts to grow at around $t=3000$ and almost saturates at about $t=7000$ as in Fig.  \ref{fig2}(a). When the density is high, the transition to the ordered state is more abrupt. Figure \ref{fig2}(b) indicates that there is an incubation time and the order parameter starts to grow at around $t=1500$ and completes within a narrow interval. 

It seems that the collective motion appears asymptotically in time even for a dilute limit since once an ordered motion appears after collisions, there is no mechanism to destroy it. If one adds noise terms in the time-evolution equations, a sharp transition from the disordered state in the dilute regime to the ordered state is expected to occur. However, we do not carry out such numerical simulations because there have been  many studies for rigid particles with noises \cite{Chate04, Mikhailov05, Perunani, Teeffelen, Chate10} and, furthermore,  the deformability is expected to be irrelevant in the dilute regime.  

The time-dependence of the order parameter in Fig. \ref{fig2}(a) can be fitted by the following function
\begin{equation}
\Phi(t)=\gamma\left[1-e^{-\alpha (t-t_0)^\beta}\right] ,
\label{avrami}
\end{equation}
where $\alpha=5.02\times 10^{-11}$, $\beta=3$, $\gamma=0.955$ and $t_0=2.25\times 10^{3}$. If $t_0$ is put to be zero, no satisfactory fit is possible.
However, the abrupt change shown in Fig. \ref{fig2}(b) cannot be represented by this type of function. This implies that the transition kinetics is quite different from the ordinary nucleation and growth in first order phase transitions in thermal equilibrium. In fact, we have seen that once an ordered region appear in the present self-propelled system, it spreads over the system almost instantaneously when the density is moderately high. 

\begin{figure}[t]
\centering
  \includegraphics[width=0.9\hsize]{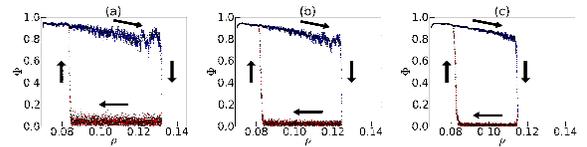}
  \caption{(Color online) Hysteresis loop 
  for (a) $N=512$, (b) $N=2048$ and (c) $N=8192$}.  
  \label{fig3}
\end{figure}

The ordered state becomes unstable for large values of density. We have investigated the disordering process by increasing the density slowly. 
The compression and expansion are made at a constant rate
\begin{equation}
 d \rho/dt=\pm \lambda,
\label{L}
\end{equation}
where 
\begin{equation}
\lambda=(\rho_{1}-\rho_{0})2^{-n},
\label{lambda}
\end{equation}
with typical values  $\rho_{0}=0.07$ and $\rho_{1}=0.14$. The exponent $n$ is varied. Figures \ref{fig3} display the order parameter $\Phi$ as a function of $\rho$ for $n=14$. It is evident that the ordered state is spontaneously broken  when the density exceeds a certain threshold value and there is a hysteresis in the transition. 
Figure \ref{fig3} shows the $N$-dependence of the hysteresis behavior. The threshold $\rho_{\mathrm{do}}$ for the transition from the disordered state to the ordered state by decreasing the density is less sensitive to the system size. We note that the threshold for the ordered-to-disordered state by increasing the density occurs at $\rho_{\mathrm{od}}\approx 0.132$ for $N=512=2^9$, $\rho_{\mathrm{od}}\approx 0.1
 25$ for $N=2048=2^{11}$ and $\rho_{\mathrm{od}}\approx 0.115$ for $N=8192=2^{13}$. This indicates that the hysteresis region $\rho_{\mathrm{od}}-\rho_{\mathrm{do}}$ has a tendency  to decrease linearly with $\ln N$. However, more accurate analysis is necessary to obtain any definite conclusion, which is left for a future study because of time consuming of numerical simulations for larger system.
We have also carried out numerical simulations for $N=512$ for slower change of the density with $n=16$ and $n=18$.  The threshold density $\rho_{\mathrm{od}}$ is approximately given by  $\rho_{\mathrm{od}}=0.132$ for $n=14$,  $\rho_{\mathrm{od}}=0.131$ for $n=16$ and  $\rho_{\mathrm{od}}=0.128$ for $n=18$. On the other hand, the threshold  $\rho_{\mathrm{do}}$  is given by  $\rho_{\mathrm{do}}=0.0840$ for $n=14$,  $\rho_{\mathrm{do}}=0.0891$ for $n=16$ and  $\rho_{\mathrm{do}}=0.0902$ for $n=18$. Therefore, as expected, the hysteresis region tends to shrink for slower density change.

\begin{figure}[t]
\centering
  \includegraphics[width=0.6\hsize]{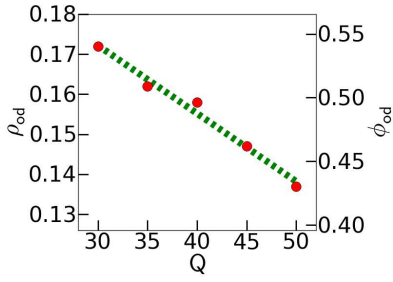}
  \caption{(Color online) The oder-disorder transition density $\rho_{\mathrm{od}}$ as a function of $Q$ for $N=512$. The dotted line is a guide to the eye. The volume fraction of the total particles defined by $\phi_{\mathrm{od}}=\pi \sigma^2 \rho_{\mathrm{od}}$ is also indicated as the scale of the vertical axis.}
  \label{fig4}
\end{figure}

The transition between the ordered and disordered states in the high density regime has not been observed in any rigid self-propelled systems and is characteristic to the deformable particles.  
The ordered state exists irrespective of the value of the anisotropic parameter $Q$ in eq. (\ref{Q}).  However,
the stability limit of the ordered state depends on it. The critical density decreases as the value of $Q$ is increased as shown in Fig. \ref{fig4}.  When the anisotropy is absent, $Q=0$, the transition to the disordered state does not occur.

We have explored the properties in the disordered state at the high density regime by evaluating numerically 
the self-intermediate scattering  function define by
\begin{equation}
F_{s}(\bm{k},t)=\frac{1}{N}\avg{\sum\limits_{j=1}^Ne^{-i\bm{k}\cdot(\bm{r}_{j}(t)-\bm{r}_{j}(0))}}.
\label{Fs}
\end{equation}
The mean square displacement of the individual particle can be obtained from $F_{s}(\bm{k},t)$ as
\begin{equation}
C(t)=\frac{\partial^2 F_{s}(\bm{k},t)}{\partial \bm{k}^2}\Big|_{\bm{k}
=0}=\avg{\abs{\bm{r}(t)-\bm{r}(0)}^{2}}=4Dt ,
\label{MSD}
\end{equation}
where $D$ is the diffusion constant in two dimensions.

\begin{figure}[t]
\centering
  \includegraphics[width=1.0\hsize]{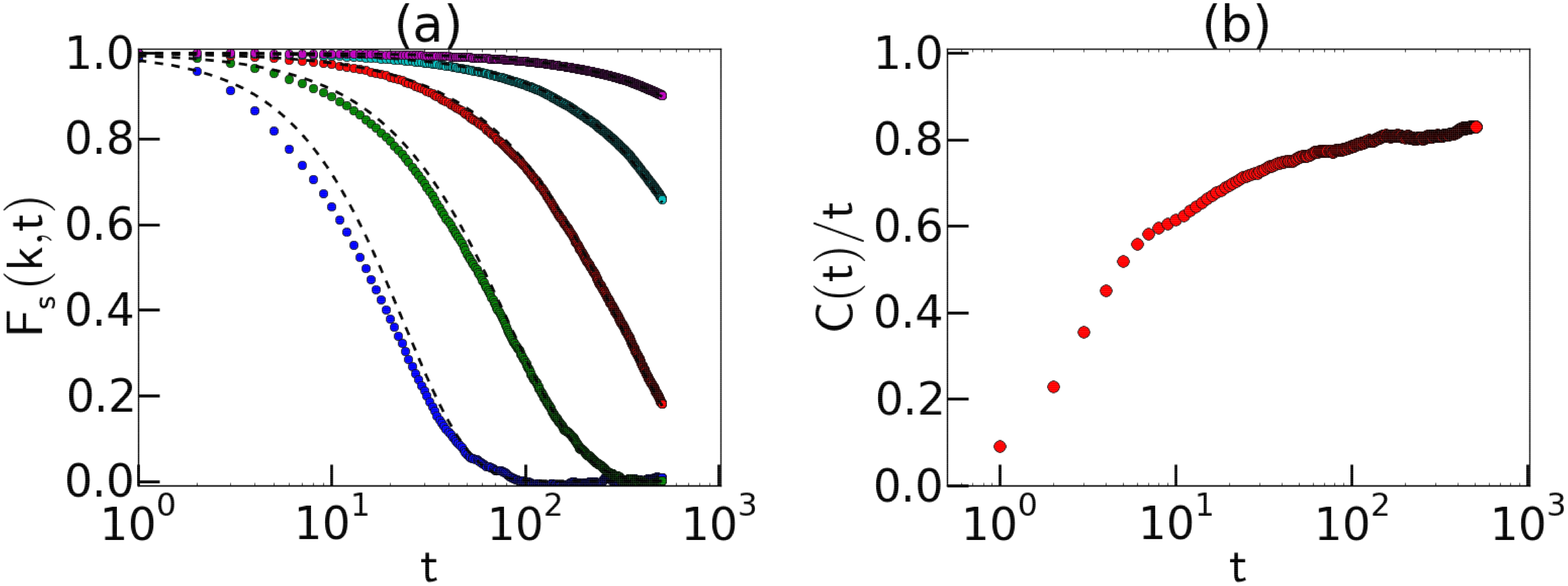} 
  \caption{(Color online) Diffusive property of the disordered state for $\rho=0.14$ and $N=512$.  (a) Self-intermediate scattering function for wave numbers $k\sigma=1/16$, $k\sigma=1/8$, $k\sigma=1/4$, $k\sigma=1/2$ and  $k\sigma=1$  from top to bottom.  The dotted curves show the function given by eq. (\ref{FS}). (b) Mean square displacement  averaged over 16 independent runs.}
  \label{fig5}
\end{figure}

The intermediate scattering function is shown as a function of time in Fig. \ref{fig5}(a) for several values of $k=\abs{\bm{k}}$. It is clear that there is no multiple relaxation which is typical in glass dynamics \cite{Binder}. Actually, as shown in Fig.  \ref{fig5}(a), the scattering function can be fitted by 
\begin{equation}
F_{s}(\bm{k},t)=\exp[-(t/\tau_k)^{\beta_k}] ,
\label{FS}
\end{equation}
where $\tau_k\approx 23.46\times k^{-\nu}$ with $\nu=1.85$.  The exponent $\beta_k$ increases from  $\beta_k\approx 1.09$ to $\beta_k\approx 1.3$ by increasing $k\sigma$ from 1/16 to 1. This indicates that the usual diffusion with the wavenumber dependence $\beta_k\nu \approx 2$ occurs for small values of $k$ whereas  a ballistic motion is dominant in the short length scale and clearly excludes the stretched exponential with $\beta_k \le 1$ \cite{Dauchot}. It is also verified (not shown) that the simple exponential decay independent of the values of $k$ is less satisfactory to fit the numerical data. 
The mean square displacement is displayed in Fig. \ref{fig5}(b). This has been  obtained by the average of 16 independent runs.
It is approximately proportional to time  asymptotically with the magnitude of $D\approx 0.2$.  
It is evident that there is no regime where $C(t)$ has a plateau due to a cage effect. This also implies that the disordered state is different from the usual glass state \cite{Binder} .

We have shown that the coherent ordered state of the deformable self-propelled particles becomes unstable
for sufficiently large density and a disordered state appears.  We have verified that the same transition occurs not only for the soft potential given by eq. (\ref{potential}) but also for 
\begin{eqnarray}
U=\exp(-r_{ij}/\sigma) . 
\end{eqnarray}
For example, the order-disorder transition occurs at around $\rho=0.02$ for decreasing the density and $\rho=0.06$ for increasing density for $N=512$ keeping all the parameters  the same values as used above.

\begin{figure}[t]
\centering
  \includegraphics[width=0.65\hsize]{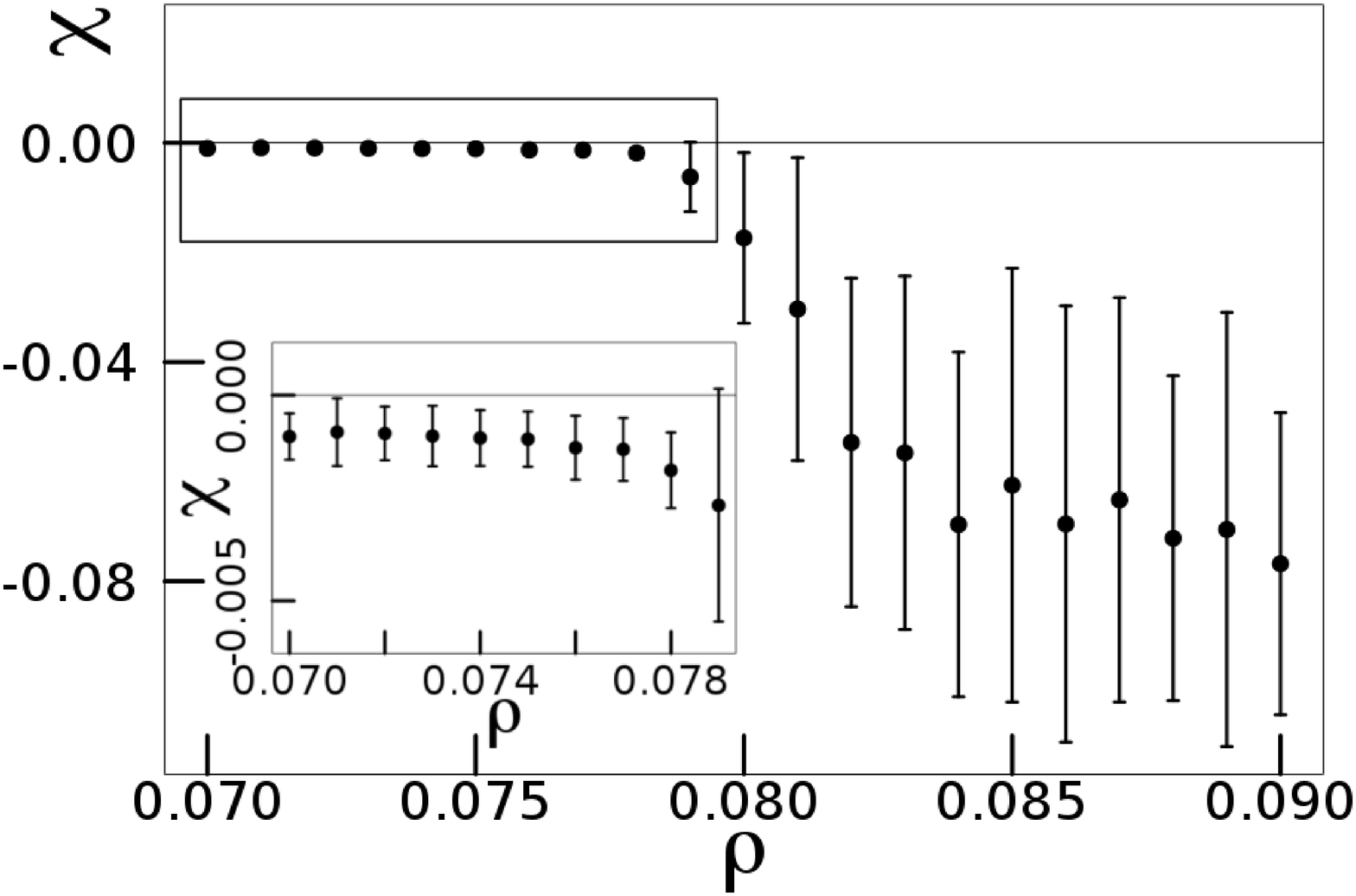} 
  \caption{Correlation $\chi$ as a function of $\rho$ for $N=512$.  The mean values are indicated by the black dots whereas the standard deviations are shown by the error bars.  This plot  is obtained by decreasing the density slowly. The transition occurs at $\rho=0.082$. The inset shows that $\chi$ is also negative for the ordered state.}
  \label{fig6}
\end{figure}

Here we describe a mean field analysis to clarify  the order-disorder transition in the high density regime. The force term in eq. (\ref{v}) can be written as
\begin{eqnarray}
 \bm{v}^{(i)}\cdot \bm{f}^{(i)}=
 K v^{(i)} \sum\limits_{j=1}^N\hat{\bm{v}}^{(i)}\cdot\bm{F}_{ij}Q_{ij} 
= Kv^{(i)} N \avg{ \hat{\bm{v}}^{(i)}\cdot\bm{F}_{ij}Q_{ij}} ,
 \label{vf} 
\end{eqnarray}
where the unit vector 
$\hat{\bm{v}}^{(i)}=\bm{v}^{(i)}/ \abs{\bm{v}^{(i)}}$ is regarded as a stochastic variable.
The average  is defined such that 
\begin{eqnarray}
 \avg{X_j}=\frac{1}{N}\sum_{j=1}^NX_j .
 \label{average} 
\end{eqnarray}
First, we make a decoupling approximation such that $Q_{ij}$ in eq. (\ref{vf}) is replaced by the average $\avg{Q_{ij}}$, 
\begin{eqnarray}
\bm{v}^{(i)}\cdot \bm{f}^{(i)}= Kv^{(i)} N \avg{ \hat{\bm{v}}^{(i)}\cdot\bm{F}_{ij}}\avg{Q_{ij}} .
 \label{vf2} 
\end{eqnarray}
By a dimensional analysis, we note that $N \avg{ \hat{\bm{v}}^{(i)}\cdot\bm{F}_{ij}}$ might have a factor $\sigma \rho$ and have verified numerically that this quantity is indeed proportional to the density $\rho$ so that we put 
\begin{eqnarray}
N \avg{ \hat{\bm{v}}^{(i)}\cdot\bm{F}_{ij}}=\rho \sigma \chi . 
 \label{chi} 
\end{eqnarray}
The unknown quantity $\chi$ is to be evaluated numerically.
Another decoupling approximation   $\avg{(S_{\alpha\beta})^2}=\avg{S_{\alpha\beta}}^2$ is also employed.
Secondly, we eliminate $S_{\alpha\beta}$ by putting $dS_{\alpha\beta}/dt=0$ in eq. (\ref{S}). By using these approximations, eq. (\ref{v}) becomes
\begin{eqnarray}
\frac{1}{2} \frac{d}{dt}w  = \gamma w -gw^{2} + \hat{\chi} \sqrt{w} [1+\hat{Q}(w-\avg{v^2})^2] ,
 \label{eqw} 
\end{eqnarray}
where $w=v^2 \ge 0$ with the superscript $(i)$ omitted and $g=1+ (ab)/(2\kappa)$,  and 
$\hat{Q}=Qb^2/(4\kappa^2)$ and
$\hat{\chi}=K\rho \chi \sigma$.
Putting $\avg{v^2} = \avg{v}^2+\avg{(\delta v)^2}$ with $\delta v$ the fluctuating part,  we identify  $\avg{v}^2$ with the stationary stable solution of eq. (\ref{eqw}). We have verified numerically that $\chi$ is negative as shown in Fig. \ref{fig6}. Therefore, we note that $w=0$ is always a stable stationary solution and that there is a pair of finite solutions, one of which is unstable and the other larger solution is stable. For sufficiently large values of $\abs{\chi}$, this pair of solutions disappear. Therefore there is a saddle-node bifurcation at some value of $\rho\equiv \rho_{\mathrm{od}}$ beyond which only the solution $w=0$ exists and hence this state corresponds to the disordered state. In the disordered branch, we should put $ \avg{v}^2=0$ in eq. (\ref{eqw}) 
whereas the stable finite stationary solution should be equated to  $\avg{v}$ in the ordered branch. 
The fact that $\chi< 0$ is essential for the saddle-node bifurcation. However, a quantitative comparison is impossible with the present crude approximation. For example, we have to require $\chi =-0.57$ ($\chi=-0.73$)  to realize $\rho_{\mathrm{od}}=0.12$ ($\rho_{\mathrm{do}}=0.082$), which are substantially different from the values in Fig. \ref{fig6}. Note also that the coefficient of the $w^{5/2}$ term which is dominant for large values of $w$ in eq. (\ref{eqw}) has a factor $\rho Q$. This implies that, if $\chi$ and $\avg{(\delta v)^2}$ are not strongly dependent on $Q$, the threshold $\rho_{\mathrm{od}}$ decreases as $Q$ is increased, which is consistent qualitatively with Fig. \ref{fig4}.
 
 We mention that the above mean field theory does not predict the intrinsic hysteresis behavior of the transition in the slow limit of the density change and the $N\to \infty$ limit. We need to take account of the stochasticity which is self-generated from the nonlinear  interaction among the self-propelled particles.
 
We emphasize that this kind of order-disorder transition has not been observed, to our knowledge, in any rigid rod-shaped particles and hence it is a characteristic feature of deformable soft particles.  It is mentioned here that nonequilibrium phase transitions due to a saddle-node bifurcation have been studied theoretically in a single-variable Langevin system \cite{Sasa}.

Finally we discuss some apparently related phenomena. 
 Colloidal particles  (although not self-propelled) exhibit glass behavior for high density. In this glass dynamics, motion of each particle is frozen or trapped in a cage and does not undergo a usual Brownian motion. This is contrast to
 the present result in Fig. \ref{fig5} where particles have a finite diffusion constant though the origin of the motion is not thermal but comes from the self-propelled energy.  A jamming transition occurs in granular materials under shear flow. The diffusion constant is finite in the jammed state but it is defined perpendicularly to the flow direction \cite{Olsson} and inherently anisotropic since the system is subjected to shear flow. %
 The third related system is a traffic flow which also exhibits a jamming state when particles are dense. However, the flow is generally directed in one dimension \cite{Cates, nishinari} or anisotropically one-dimensional even in the two-dimensional space \cite{Biham} and therefore the jammed state has no direct analogue with the present isotropic disordered state. 

\section*{Acknowledgements} 
This work was supported by the Grant-in-Aid for the priority area ``Soft Matter Physics''
 and the Global COE Program ``The Next Generation of Physics, Spun from Universality and Emergence''
 both from the Ministry of Education, Culture, Sports, Science and Technology (MEXT) of Japan.

\end{document}